\documentstyle[11pt,newpasp,twoside]{article}
\def\edcomment#1{\iffalse\marginpar{\raggedright\sl#1\/}\else\relax\fi}
\reversemarginpar

\begin{document}
\title{The Hydrogen Clouds in the Galactic Halo}
 \author{Felix J. Lockman}
\affil{National Radio Astronomy Observatory, Green Bank WV, USA}

\begin{abstract}
New 21cm observations with the Green Bank Telescope (GBT) 
show that a significant fraction of the 
HI in the inner Galaxy's halo $\sim 1$ kpc from the midplane exists 
in the form of discrete clouds.  
Some look very much like a Spitzer (1968) ``standard'' 
diffuse cloud but with their HI in two phases.  
They mark the transition between 
the neutral disk and the highly ionized halo.
The dominant motion of the clouds is Galactic rotation, but 
some have  random velocities of as much as 50 km~s$^{-1}$.  They are 
part of the Galaxy and are not related to high-velocity clouds, yet their 
origin is obscure.  
\end{abstract}

\section{Introduction}

In the inner parts of the Milky Way most of the hydrogen is confined to a 
narrow disk which establishes a plane of fundamental physical 
significance (Blaauw et al. 1960).   But the more we look out of the disk, 
the more HI we see at curious locations.  
Many mechanisms  can lift HI from below, and 
the Milky Way, like other galaxies,  is probably accreting 
 gas from above, so the halo is likely an untidy place, 
containing HI  that  1) has been pushed outward by mechanical and photon 
energy from young stars and supernovae, 2)  has condensed 
from a halo of very hot gas and now returns in a cool phase to the plane, 
3) is falling into the Milky Way for the first time, having been 
 stripped from neighboring galaxies like the Magellanic Stream, or  
coming in as  high-velocity clouds (Shapiro \& Field 1976;  
Heiles 1984; Norman \& Ikeuchi 1989; Houck \& Bregman 1990; 
Wakker et al. 1999; Sancisi et al. 2001; de Avillez \& Berry 2001; 
Konz et al. 2002; Putman et al. 2003).  

This paper discusses new observations of the HI at some distance from 
the Galactic plane which show the organization of halo HI for the 
first time, but at present raise more questions than answers.

\begin{figure}
\plotfiddle{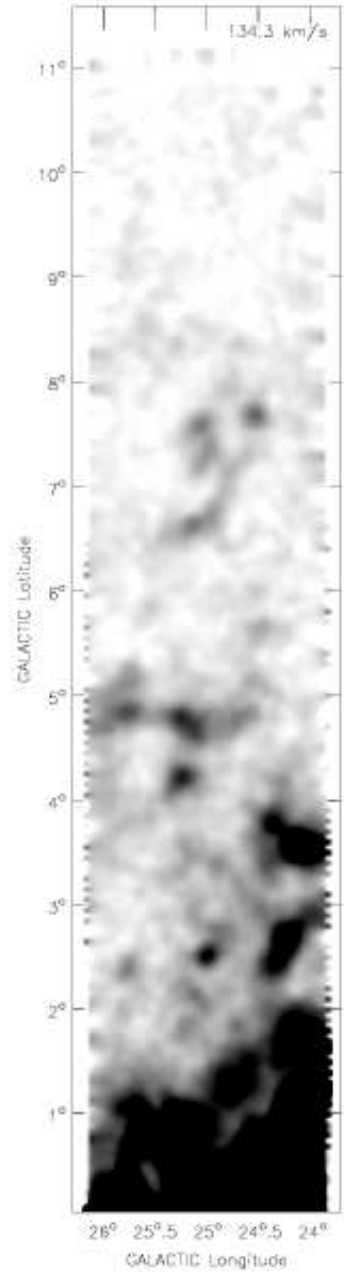}{6.45in}{0}{80}{80}{-280}{-75}
\caption{HI emission above the Galactic plane in a region around 
 $\ell = 25\deg$ as observed with the GBT at $9'$ angular 
resolution.  The data cover  $V_{LSR} = 134 \pm 1$ 
km~s$^{-1}$, which is at the tangent 
point in this direction, where one degree in either coordinate 
corresponds to a linear distance of 135 pc.  The clouds at 
$b \approx +7\fdg6$ are 1 kpc above the plane. 
}
\end{figure}

\section{The Structure of HI Above the Disk}

Previous studies of 21cm emission in the inner Galaxy were able to 
detect neutral gas to considerable distances from the Galactic plane 
(Oort 1961; Lockman 1984), but lacked  the angular resolution 
to determine its structure.  That situation has now changed 
because of the  
 Green Bank Telescope (GBT), a 100-meter diameter offset paraboloid 
antenna,  which has very good sensitivity, high dynamic range, 
 and an angular resolution of about $9'$ in the 21cm line.  

Figure 1 shows recent GBT observations of 
 HI above the Galactic plane around $\ell = 25\deg$. 
There is some HI which might be a diffuse component 
 falling off gradually with 
latitude, but the most striking features  are the  clouds which stand out 
 above the background.  These clouds are 
 well-defined in position and velocity, are several 
orders of magnitude denser than their surroundings, and may contain a 
significant fraction of the neutral gas in the halo.
Similar clouds are seen 
throughout the inner Galaxy (Lockman 2002).  
The halo clouds follow Galactic rotation, so their origin is 
tightly tied to events in the disk and lower halo.  In the rest 
of this paper I will discuss a few features of this population  
 and give provisional answers to some of the most obvious 
questions.

\section{A Halo Cloud and its Surroundings}

Figure 2 shows a series of GBT HI spectra across the  halo cloud 
at $25\fdg1+7\fdg6$ from Fig.~1. 
The spectra show: 1) this cloud is not 
sitting on a plateau of HI emission at the same velocity ---  
it contains most of the HI emission at its velocity in its vicinity.
2) At the edge of the cloud the line shape changes.  
Closer inspection shows that 
this cloud has two line components, one with a FWHM 
 of 8 km~s$^{-1}$, and the other, at an identical velocity, with 
a FWHM of 26 km~s$^{-1}$.  The narrower line is confined to the center 
of the cloud.  This cloud 
(and many others) thus contains HI in at least 
two phases, a condition which is observed in other 
locations (Liszt 1983), and  which can occur only over a 
restricted range of physical conditions (Field, Goldsmith \& Habing 
1969; Wolfire et al. 1995). 
 3)~This cloud appears to be connected to several others by 
a narrow filament which has a linewidth similar to
 the broad component of the cloud.  Such connections are also seen  
between a few other clouds, but not all.   The cloud 
has a mass in HI $ \approx 400 M_{\circ}$, a size of $50 \times 35$ pc, 
and a peak $N_H \geq 3 \times 10^{19}$  cm$^{-2}$.

\begin{figure}[ht]
\plotfiddle{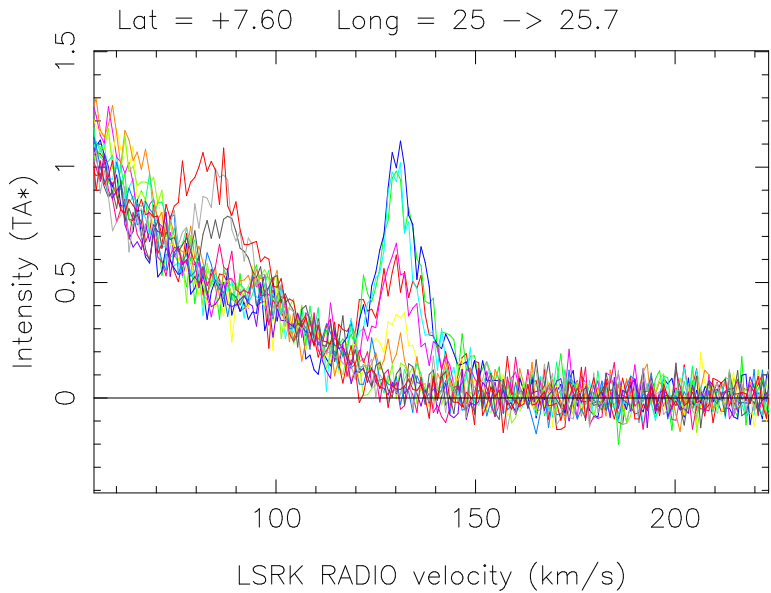}{3.in}{-90}{45}{45}{-200}{255}
\caption{GBT 21cm spectra across the halo cloud at latitude $7\fdg6$ 
between $25\deg \leq \ell \leq 25\fdg7$ at $3'$ spacing.  There 
is little emission away from the cloud.  The line changes shape 
at the cloud edge.
}
\end{figure}

\section{The Velocity of the Clouds}

A set of about  40 halo clouds near longitude $29\deg$, 
all more than 500 pc from the plane, 
 were studied in some detail (Lockman 2002).  Here I will 
 concentrate on their velocities. 
  Fig.~3 shows the number of clouds  in  intervals 
of $V_{LSR}$.  The distribution implies a 
cloud-to-cloud velocity dispersion of several 
tens of km~s$^{-1}$.  Verification of this will require more careful 
analysis, for at the lower velocities clouds are so common that 
they are confused and probably under-counted.  Nonetheless, some 
of the ``fastest'' cloud have random velocities of $> 40$ km~s$^{-1}$. 
The fastest clouds do not have a preferential location with 
respect to the plane, but seem to be found at all latitudes. 
    The only property of the clouds 
which is correlated with  velocity is  mass: the higher 
the random velocity the smaller the mass.  Thus,  in the halo the  HI mass 
is concentrated toward velocities permitted by Galactic rotation, 
just as it is in the disk.  

It was suggested some time ago that the Galaxy contained
a population of  ``fast'' interstellar clouds 
whose random velocities might carry them into the halo 
(Radhakrishnan \& Srinivasan 1980; Kulkarni \& Fich 1985; 
Lockman \& Gehman 1991), but 
individual clouds could not be resolved in the older data. 
Now we know of distinct HI clouds with random velocities of 
several tens of km~s$^{-1}$ not only in the halo, but at 
low latitude as well (Lockman \& Stil 2004).  This 
population may be quite widespread in the inner 
Galaxy, and  may also be linked to the 
broad HI profile components found by Kalberla et al. (1998).  

\begin{figure}
\plotfiddle{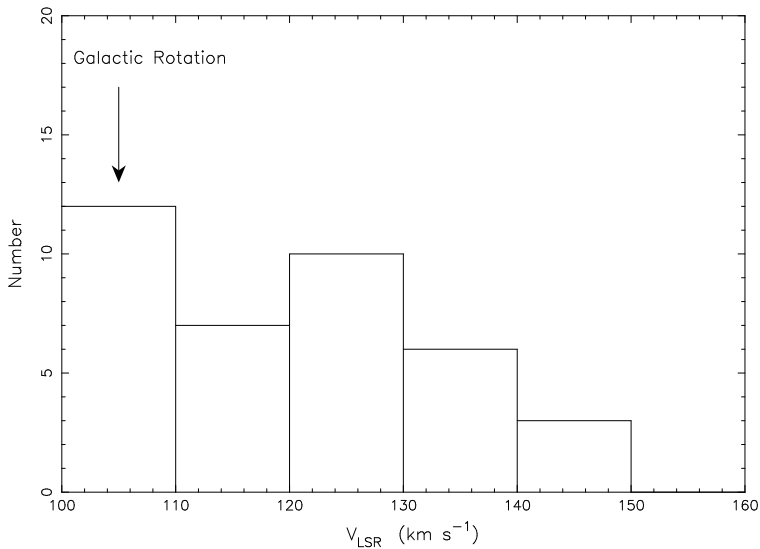}{2.9in}{-90}{40}{40}{-170}{240}
\caption{The distribution of $V_{LSR}$ for halo HI clouds in a sample  
near $\ell = 29\deg$ (Lockman 2002).  These clouds are 
at a  median distance of --940 pc from the plane.  The arrow 
marks the maximum velocity expected for 
 Galactic rotation in this direction, as 
determined from the velocity of molecular clouds in the disk 
(Clemens 1985).
}
\end{figure}

\section{Comments on the Origin of the Halo Clouds}

Because the halo clouds  are firmly tied to Galactic rotation 
even though some have a large random velocity, 
they cannot be extragalactic material falling into the disk, like 
some  high-velocity clouds  (e.g., Wakker et al. 1999).  
So where did the halo clouds come from?  Possible sources were 
discussed in $\S1$: all involve massive stars and 
supernovae which can push 
HI outward and also create hot gas for a Galactic ``fountain''. 
There is ample evidence that these processes are at work 
(e.g., Savage 1995; Sembach et al. 2003), but 
it is not obvious to me that the basic theoretical models have been 
 developed far enough to be tested against the new data.

The clouds detected so far look approximately spheroidal, even 
when observed at $1'$ angular resolution (Liszt, Lockman \& Rupen, 
in preparation).  They look stable, though their dynamical time 
(size/linewidth) is only a few $10^6$ years.   
Does something surround the clouds keeping them in equilibrium? 
  At a distance of 1 kpc from the 
Galactic plane the average HI density in the inner Galaxy is 
expected to be 0.005 cm$^{-3}$ (Dickey \& Lockman 1990), while 
the average density of H$^+$ will be about twice this value,  
though it probably fills only a fraction of the volume (Reynolds 
1997).  A typical  halo cloud has $\langle n \rangle \approx 0.25$ cm$^{-3}$ 
(Lockman 2002) suggesting that if most of the halo HI is in 
clouds, the filling factor of neutral gas 
is only a few percent.  The remainder of the space may be 
 filled with very hot gas which 
provides confinement of the halo clouds if, indeed, 
they are pressure-confined, but this an assumption not an 
observation (e.g., Cox 2000).  One certainty is that 
the clouds have too little HI mass to be self-gravitating.  

Whatever holds them together, at 1 kpc height 
the halo clouds are denser than their surroundings by more than an 
order of magnitude and left to themselves will plunge 
toward the plane.  It is plausible that they have a vertical velocity 
of at least the same magnitude as their random line-of-sight velocity, i.e., 
 several tens of km~s$^{-1}$, though whether
the dominant motion is toward the midplane as would be expected from 
fountain models, or is in equilibrium (some clouds rising, others falling), 
cannot be resolved by the data in hand.

\section{Prospects}

The study of halo clouds is just beginning.  
They are likely to be
sensitive tracers of physical conditions in their vicinity and 
could be used as probes of their environment.  Moreover, 
they are the best examples known of diffuse interstellar clouds of the kind 
envisioned by Spitzer (1968), though they lie in the halo, not the disk. 
In fact, the halo clouds are the {\it only} diffuse interstellar 
clouds which show well-defined boundaries in position and velocity 
when mapped in HI  (cf., Kulkarni \& Heiles 1988). 
  They may be useful not only for study of 
conditions in the halo, but for study of fundamental  interstellar 
processes.  The GBT data make it clear that the 
neutral halo has more structure, and is more 
amenable to analysis, than had been suspected before.  

\acknowledgments{The National Radio Astronomy Observatory is
 operated by Associated Universities, Inc., under a cooperative 
agreement with the National Science Foundation.}

\end{document}